\documentclass[aps,prd,preprint,showpacs,preprintnumbers,nofootinbib]{revtex4}
\usepackage{latexsym,amsmath,amssymb,amstext}
\usepackage{float}
\usepackage{graphicx,xcolor}
\usepackage[paperwidth=210mm,paperheight=297mm,centering,hmargin=2cm,vmargin=2.5cm]{geometry}

\newcommand{\be}{\begin{equation}}
\newcommand{\ee}{\end{equation}}
\newcommand{\bea}{\begin{eqnarray}}
\newcommand{\eea}{\end{eqnarray}}

\newcommand\bef{\begin{figure}}
\newcommand\eef[1]{\label{fg:#1}\end{figure}}
\newcommand\beq{\begin{equation}}
\newcommand\eeq[1]{\label{#1}\end{equation}}
\newcommand\beqa{\begin{eqnarray}}
\newcommand\eeqa[1]{\label{#1}\end{eqnarray}}
\newcommand\bet{\begin{table}}
\newcommand\eet[1]{\label{tb:#1}\end{table}}

\newcommand\fgn[1]{Figure~\ref{fg:#1}}
\newcommand\eqn[1]{Eq.~(\ref{#1})}

\newcommand\tbn[1]{Table~\ref{tb:#1}}

\begin{document}
\title{Bilinear condensate in three-dimensional large-$N_c$ QCD}
\author{Nikhil\ \surname{Karthik}}
\email{nkarthik@fiu.edu}
\affiliation{Department of Physics, Florida International University, Miami, FL 33199.}
\author{Rajamani\ \surname{Narayanan}}
\email{rajamani.narayanan@fiu.edu}
\affiliation{Department of Physics, Florida International University, Miami, FL 33199.}

\begin{abstract}
We find clear numerical evidence for a bilinear condensate in
three-dimensional QCD in the 't~Hooft limit.  We use a non-chiral
random matrix model to extract the value of the condensate $\Sigma$
from the low-lying eigenvalues of the massless anti-Hermitian overlap
Dirac operator.  We estimate $\Sigma/\lambda^2 = 0.0042\pm 0.0004$
in units of the physical  't~Hooft coupling.
\end{abstract}

\date{\today}
%\pacs{}
\maketitle

It has been recently shown~\cite{Karthik:2015sgq}, contrary to
expectations (see references in~\cite{Karthik:2015sgq}), that QED
in three dimensions with parity-invariant coupling to massless
two-component fermions does not result in a bilinear condensate for
any number of fermion flavors.  Further numerical analysis using
overlap fermions~\cite{Karthik:2016ppr} that preserves the U$(2N_f)$
global symmetry of the theory on the lattice with $2N_f$ flavors
of two-component fermions shows that the theory behaves in a
scale-invariant manner.

Three-dimensional QCD with SU$(N_c)$ as the color group and $2N_f$
flavors of two-component fermions has also been studied with the
aim of finding a critical number of fermion flavors below which the
theory is confined and develops bilinear condensate for massless
fermions.  Analysis of the gap equation~\cite{Appelquist:1989tc}
suggests the existence of such a critical number of fermion flavors.
A numerical study of the SU$(3)$ gauge theory in the quenched
approximation using staggered fermions has shown evidence for a
bilinear condensate at finite lattice spacings~\cite{Damgaard:1998yv}.
This issue has been recently studied in~\cite{Goldman:2016wwk} using
the $\epsilon$-expansion about four dimensions.

Consider the theory in the limit of $N_c\to\infty$.  If the fermions
are coupled in a parity-invariant manner, then the fermion determinant
is real and positive, and does not contribute to the measure in the
$N_c\to\infty$ limit, provided $N_f$ is kept
finite~\cite{'tHooft:1973jz,'tHooft:1983wm}. The pure gauge theory
in the large-$N_c$ limit is in the confined phase at zero temperature,
and undergoes deconfinement transition at a temperature $T_c$.  The
continuum reduction~\cite{Narayanan:2007ug} implies that the theory
in a periodic box of size $\ell_x\times \ell_y \times \ell_z$ is
in the confined phase if $\ell_x,\ell_y,\ell_z > \frac{1}{T_c}$ and
there is no dependence on the box size.  A computation of the string
tension using a variational technique~\cite{Karabali:1998yq}; a
numerical evaluation~\cite{Bringoltz:2006zg} at finite values of
$N_c$ extrapolated to $N_c\to\infty$; and a numerical
evaluation~\cite{Kiskis:2008ah} at large $N_c$ using continuum
reduction are all in good agreement with each other. Since fermions
do not provide a back reaction in the 't~Hooft limit and the theory
has a non-zero string tension, we expect massless fermions to develop
a non-zero bilinear condensate in the large-$N_c$ limit. A numerical
study establishing the presence of a bilinear condensate using
techniques similar to the ones used in~\cite{Karthik:2016ppr} will
serve as a sanity check and justify a future numerical study of
SU$(N_c)$ gauge theory coupled to $2N_f$ flavors of dynamical
fermions and map the critical line in the $(N_c,N_f)$ plane that
separates the phase where scale invariance is broken from one that
is scale-invariant.  This is the aim of this brief report.

We used the standard Wilson gauge action and $b$ is the lattice
gauge coupling, which is related to the physical 't~Hooft coupling,
$\lambda=g^2N_c$, by
\be
b=\frac{1}{a \lambda},
\ee
where $a$ is the lattice spacing.  We used the primes
$N_c=7,11,13,17,19,23,29,37,41$ and 47 in this study. We worked on
a periodic $L^3$ lattice. Based on the numerical studies
in~\cite{Narayanan:2007ug}, we know that $b\in [0.55,0.75]$ is in
the confined phase as long as $L \ge 4$.  We used five different
lattice couplings; $b=$ 0.55, 0.6, 0.65, 0.7 and 0.75 on $4^3$
lattice; to study the approach to the continuum limit.  We also
used $L=4,5$ and 6 at $b=0.75$  to check for any volume dependence.
We used overlap fermions with the standard Wilson kernel as described
in~\cite{Karthik:2016ppr} and studied the behavior of the five
low-lying eigenvalues. We used the Cabibo-Marinari SU$(2)$ heat
bath along with the SU$(N_c)$ over-relaxation
algorithm~\cite{Kiskis:2003rd} to generate 300-500 statistically
independent gauge field configurations for the pure gauge theory.
Details pertaining to the overlap Dirac operator in three dimensions
and the computation of low-lying eigenvalues can be found
in~\cite{Karthik:2016ppr}.

\bef
\centering

\includegraphics[scale=0.75]{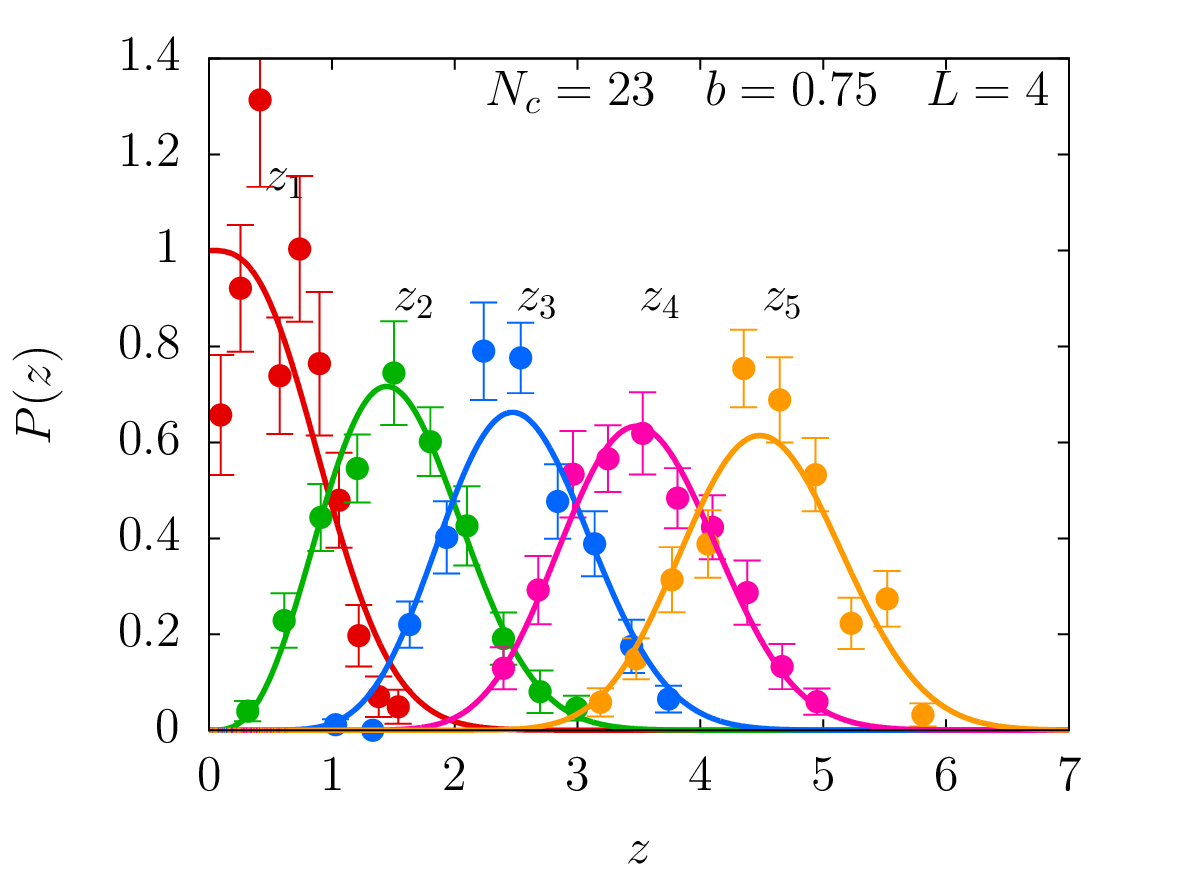}

\includegraphics[scale=0.75]{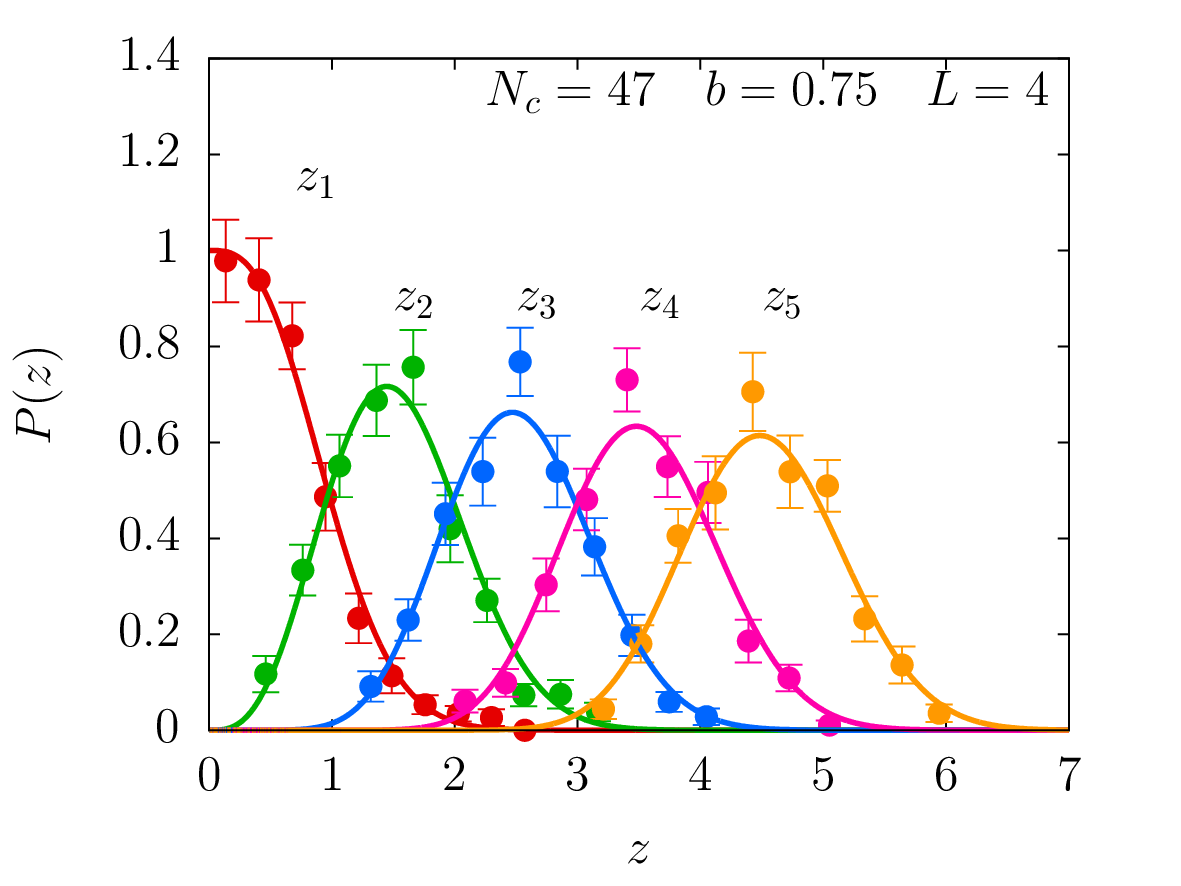}

\caption{The agreement between the distributions of the five scaled
low-lying eigenvalues (data points) of the overlap Dirac operator,
$N_c L^3 \Sigma(i,N_c,L)\Lambda_i$, and the distributions from the
non-chiral random matrix model (solid curves) is shown. All the
data are on $4^3$ lattice.  Two different $N_c$'s are shown~:
$N_c=23$ on the top panel and $N_c=47$ on the bottom.  Agreement
with the non-chiral RMM gets better when $N_c$ is increased.  }
\eef{histo}

\bef
\centering

\includegraphics[scale=0.75]{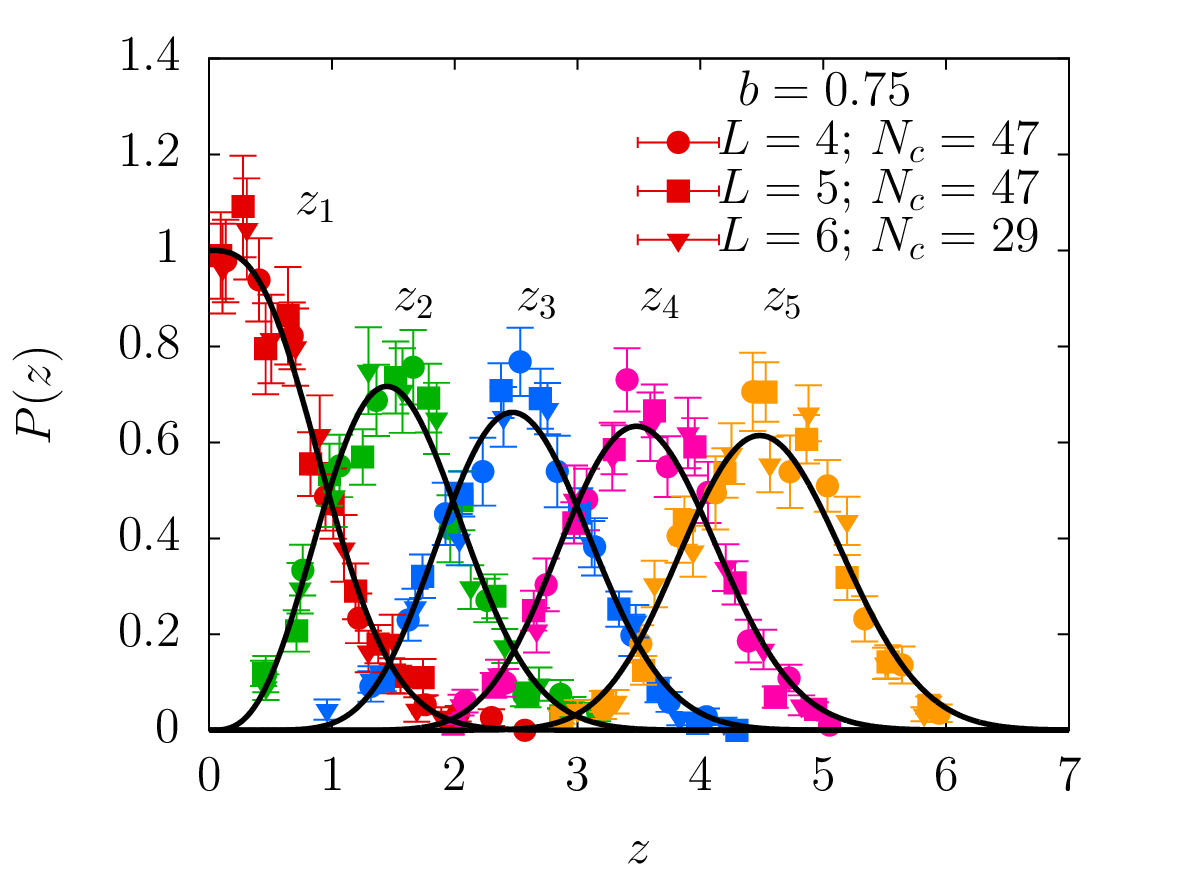}

\includegraphics[scale=0.75]{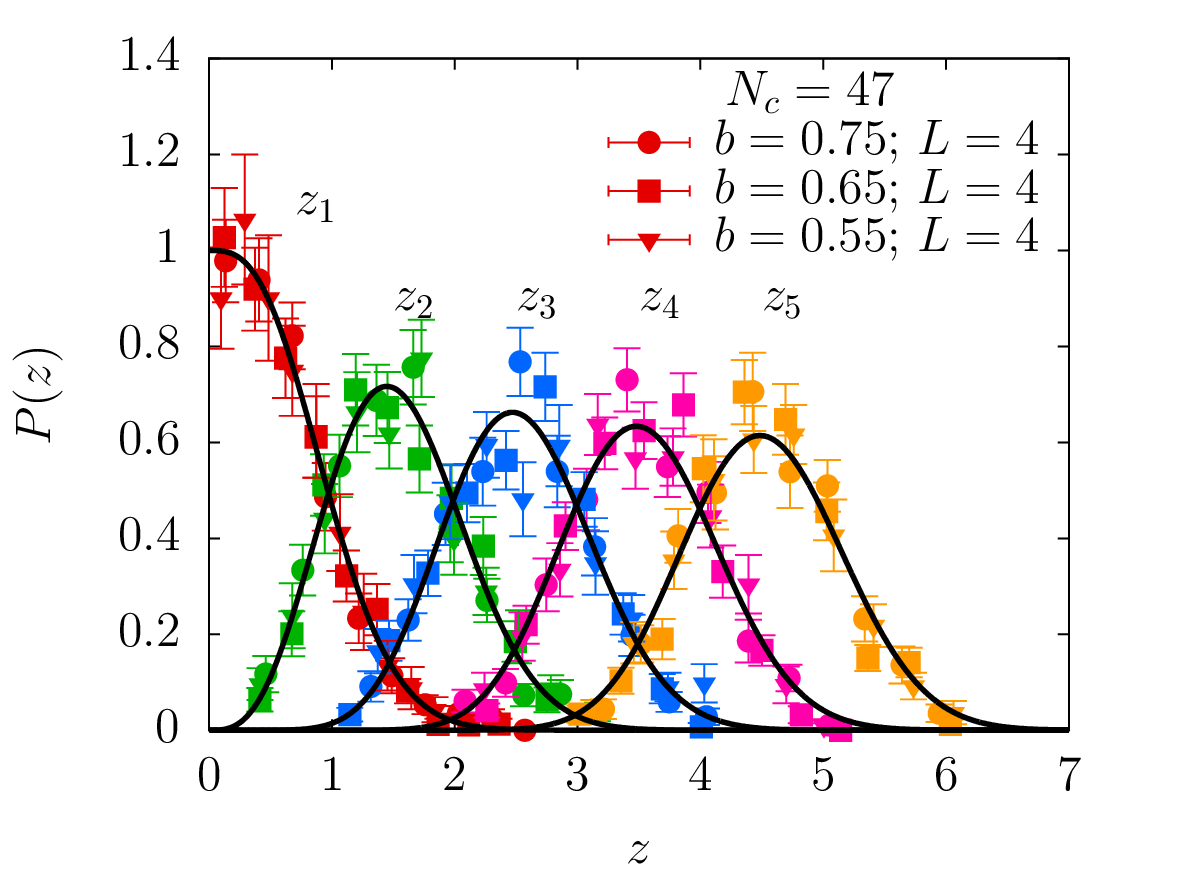}

\caption{ On the top panel, the distributions of the scaled eigenvalues
of the overlap Dirac operator at various lattice sizes $L=$ 4, 5
and 6, at the same $b=0.75$, are compared with the
distributions from the non-chiral RMM. An agreement is seen independent
of the volume. On the bottom panel, such a comparison between the
data at $b=$ 0.75, 0.65, 0.55  and the non-chiral RMM distributions
is made at the same $L=4$ and $N_c=47$.  The agreement is seen at
all lattice spacings in this study. }
\eef{bldep}

The eigenvalues $i\Lambda_j$ are associated with an anti-Hermitian
operator in the case of overlap fermions. There is no symmetry in
three dimensions that pairs up eigenvalues of opposite signs per
configuration. The parity symmetry implies that the spectrum is
flipped about zero under parity.  Therefore, the distribution of
eigenvalues will be symmetric around zero. The presence of a bilinear
condensate implies a non-zero density at zero eigenvalue. Level
repulsion implies that the level spacing of eigenvalues near zero
will be inversely proportional to $N_c L^3$. The individual
distributions of the low-lying eigenvalues (ordered by their absolute
values) will be governed by an appropriate non-chiral random matrix
model (RMM)~\cite{Verbaarschot:1994ip,Szabo:2000qq}, which in our
case will be a Hermitian random matrix model: the matrix elements
of a $k\times k$ Hermitian matrix, $H$, are independently and
normally distributed with zero mean and a variance of $\pi^2/4k$.
The spectrum of each randomly generated $H$ will not be symmetric
about zero but the distribution will be symmetric on the average
since $H$ and $-H$ have the same weight.  The distributions of the
low lying eigenvalues $z_j$ in the RMM model can be obtained using
the sinc-kernel and the associated Fredholm
determinants~\cite{Mehta:2004,Nishigaki:2016nka}.  We numerically
evaluated the eigenvalues of the kernel required for the computation
of the determinants and traces of the resolvents, and we were able
to determine the distributions of the five lowest eigenvalues $z_j$
in the RMM needed for our comparison to a very good accuracy.

The bilinear condensate can be obtained by matching the distribution
in the large-$N_c$ gauge theory to the RMM model in the large $k$
limit.  In theory, for very large $N_c$ one should be able to make
such a matching for all the eigenvalues using a single number
$\Sigma_{\rm lat}(b)$, which is the condensate. In practice, at
finite $N_c$ we scale the $j$-th eigenvalue by $\Sigma_{\rm
lat}(j,N_c,b,L)$ such that their respective distributions $P_j$
match:
\be
P_j\big(\left\{ N_c L^3 \Sigma_{\rm lat} (j,N_c,b,L)\big\} \Lambda_j\right) = P_j\left(z_j\right),
\ee
where $i\Lambda_j$ is the $j$-th eigenvalue of the anti-Hermitian
overlap Dirac operator computed in the quenched SU$(N_c)$ gauge
theory on a $L^3$ lattice at lattice gauge coupling $b$, and $z_j$
is the $j$-th eigenvalue of $H$ in the $k\to\infty$ limit.
If a non-zero condensate $\Sigma_{\rm lat}(b)$ is present
in the large-$N_c$ theory on the lattice, then
\be
\Sigma_{\rm lat}(b) = \lim_{N_c\to\infty} \Sigma_{\rm lat}(j,N_c,b,L) \ne 0,
\ee
and it should be independent of  $j$ (only one scale parameter) and
$L$ (lattice volume independence) for large enough $N_c$.  If a
non-zero condensate $\Sigma$ is present in the continuum limit of
the large-$N_c$ theory, then
\be
\frac{\Sigma}{ \lambda^2} = \lim_{b\to\infty} \Sigma_{\rm lat}(b) b^2.
\ee
With the intention of obtaining the continuum limit, we consider
the quantity, $b^2 \Sigma_{\rm lat}(b)$, in the following discussion.
In \fgn{histo}, we make a comparison of distributions at two different
values of  $N_c$ ($=23$ and $47$), at the finest lattice spacing
used in this study.  An agreement between the scaled eigenvalues
of the overlap operator, and the non-chiral RMM distributions is
seen for the low-lying eigenvalues.  As one would expect in the
presence of a bilinear condensate, this agreement is seen to get
better as $N_c$ is made larger. Further, we find this agreement
with the non-chiral RMM for three different lattice volumes at a
fixed lattice coupling as shown in the top panel of \fgn{bldep}.
The agreement with RMM continues to hold as one changes the lattice
coupling as seen in the bottom panel of \fgn{bldep}.

\bef
\centering
\includegraphics[scale=0.68]{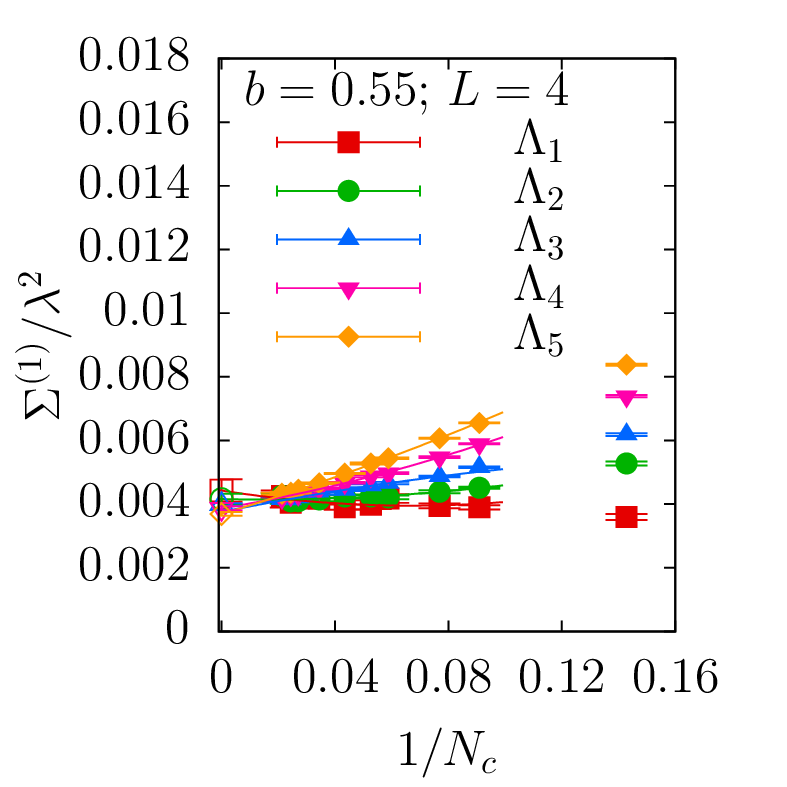}
\includegraphics[scale=0.68]{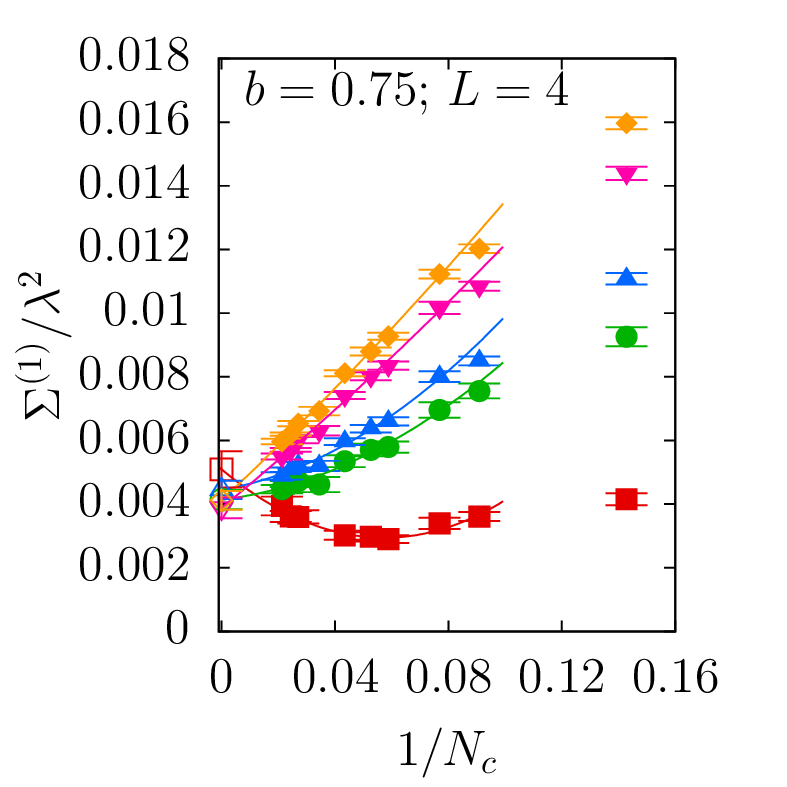}
\includegraphics[scale=0.68]{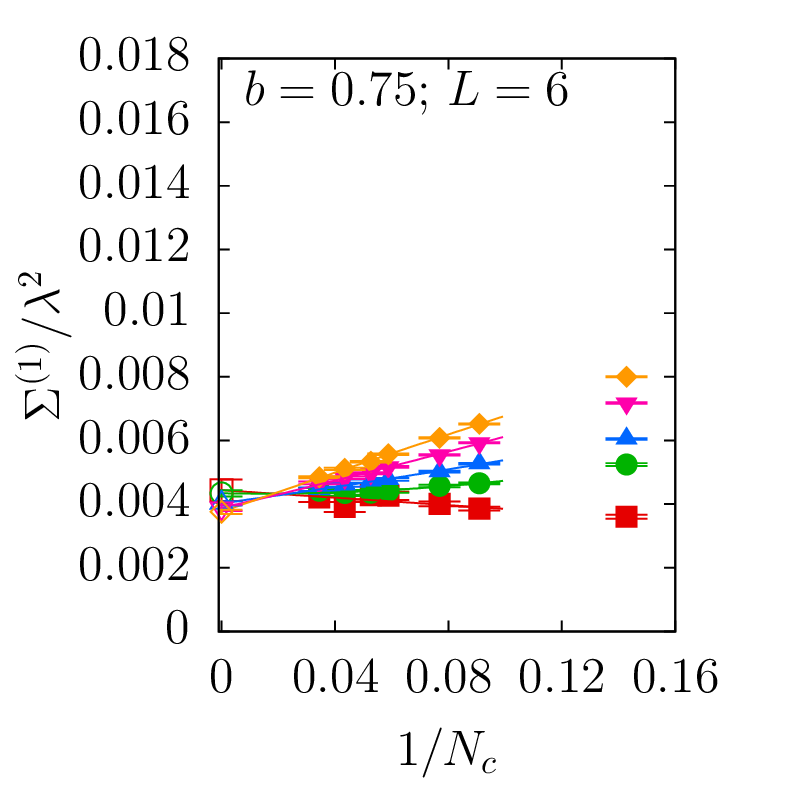}

\includegraphics[scale=0.68]{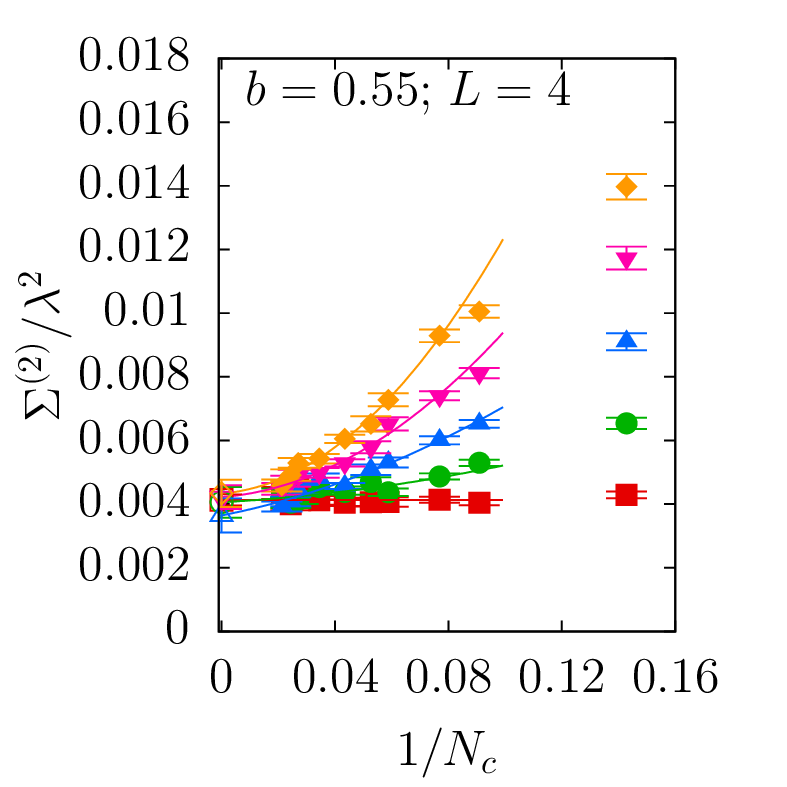}
\includegraphics[scale=0.68]{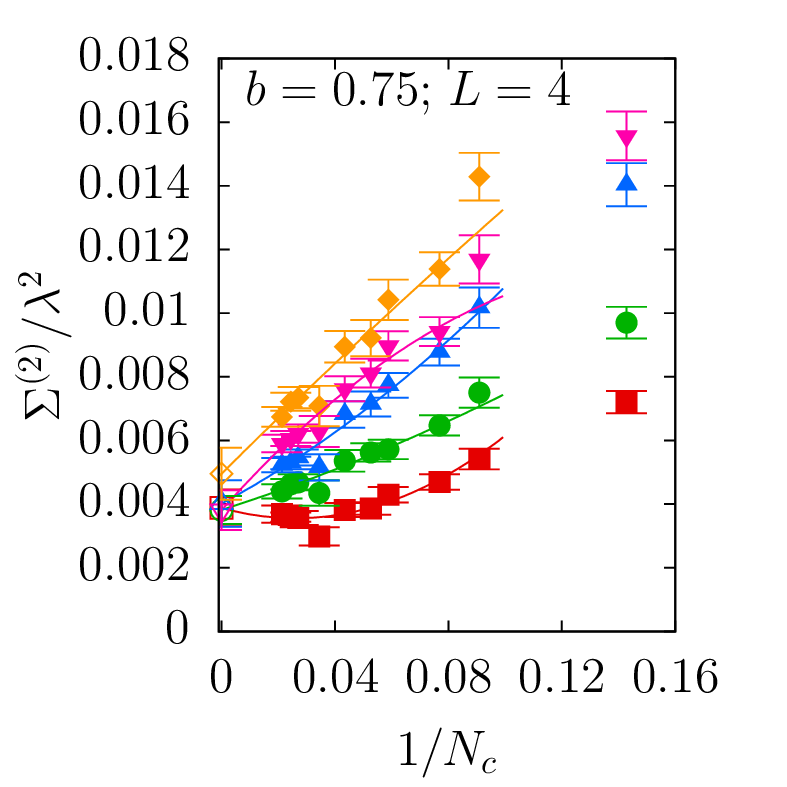}
\includegraphics[scale=0.68]{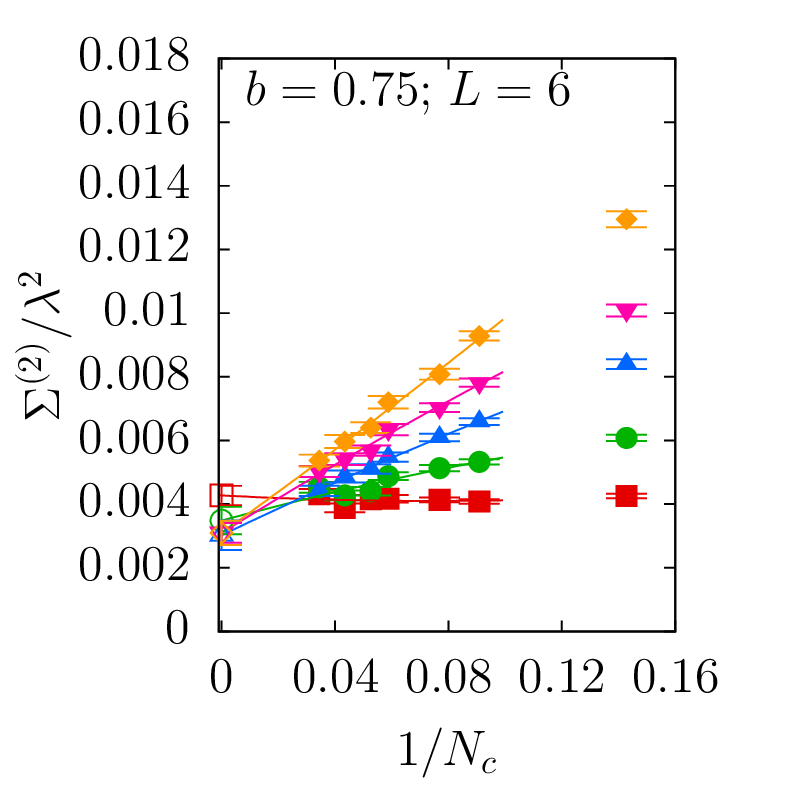}
\caption{Infinite $N_c$ extrapolation of $\Sigma/\lambda^2$.  The
top panels are for $\Sigma^{(1)}/\lambda^2$ estimated from the means
of the first five low-lying eigenvalues of the overlap Dirac operator,
while the bottom ones are for $\Sigma^{(2)}/\lambda^2$ estimated
from standard deviation of the distributions (refer
\eqn{sigdef}).  The leftmost panels are at a lattice
coupling $b=0.55$ on $L=4$ lattice.  The center panels are at
$b=0.75$ on $L=4$ lattice. The rightmost panels are at the same
$b=0.75$ but on $L=6$ lattice.  }
\eef{largen}

\bef
\centering
\includegraphics[scale=0.75]{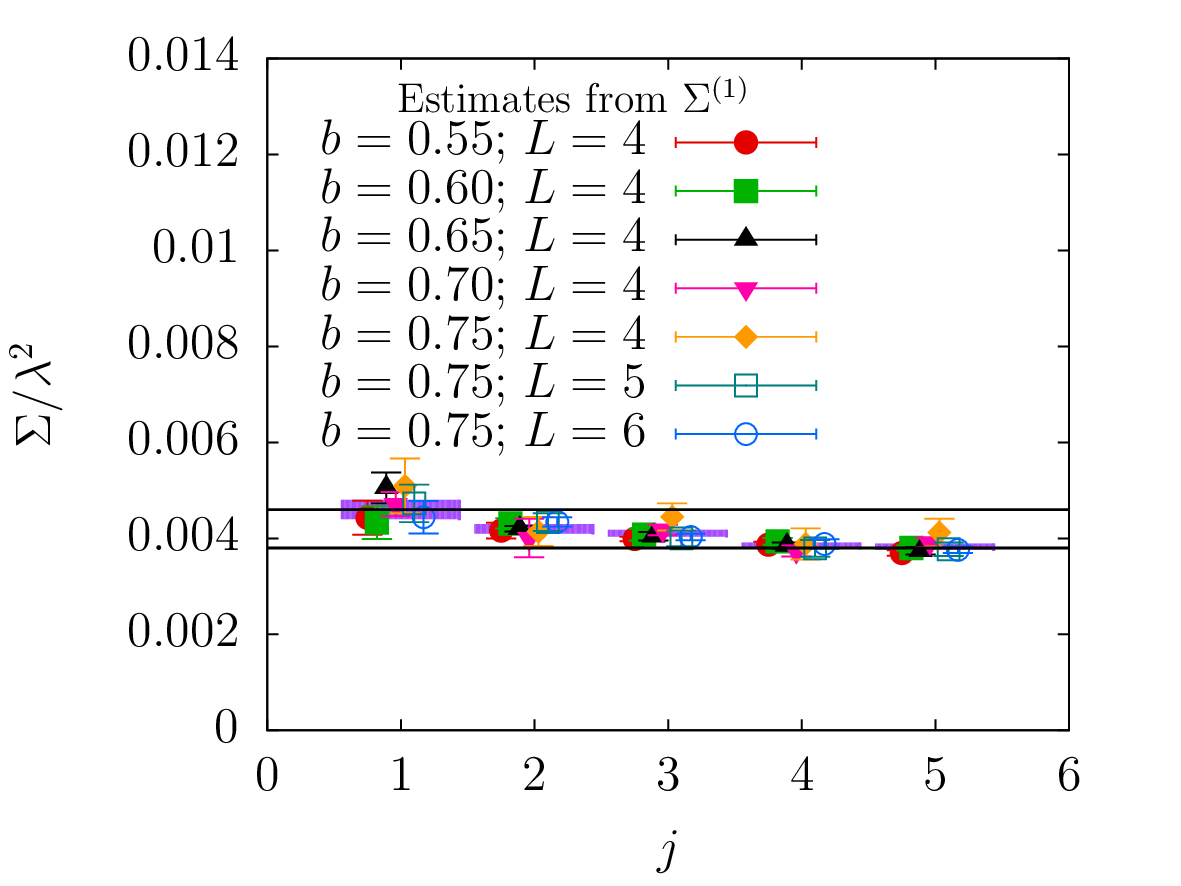}

\includegraphics[scale=0.75]{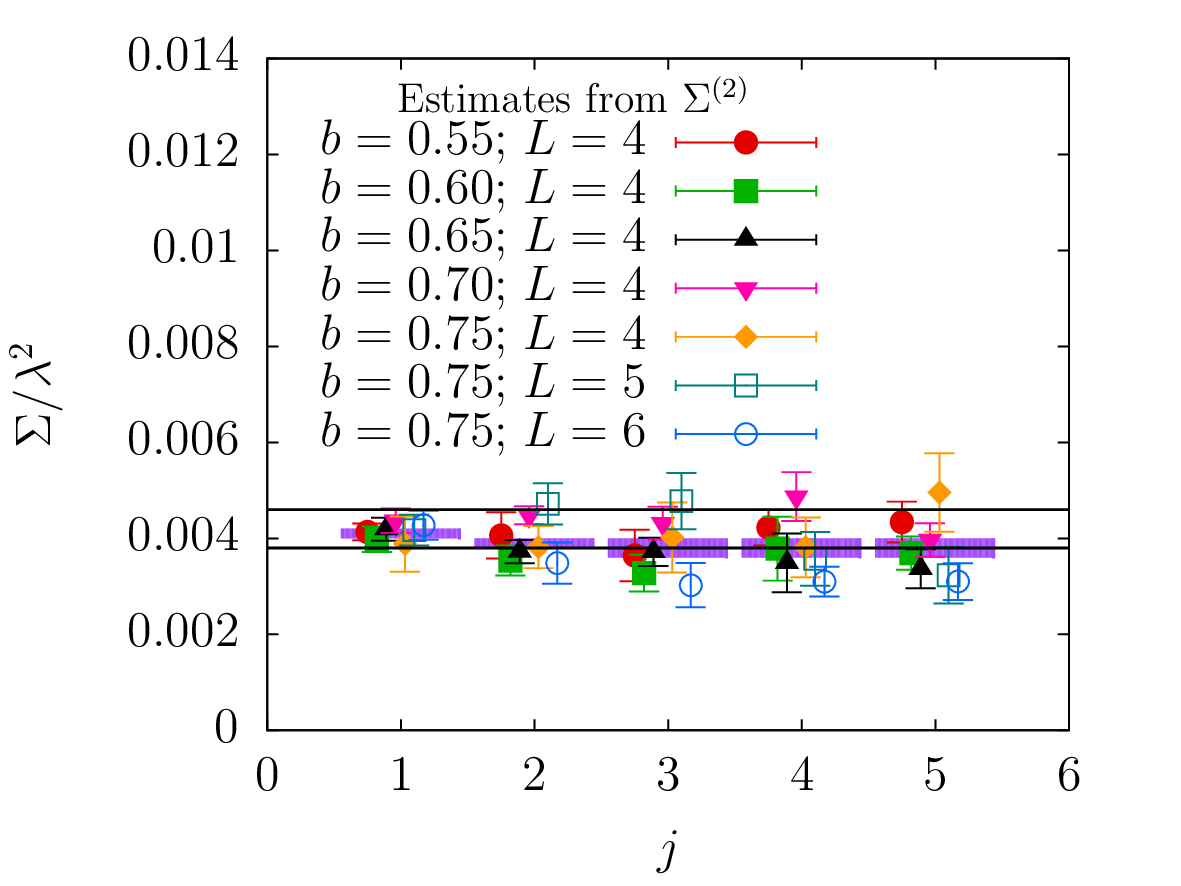}
\caption{
Bilinear condensate at infinite $N_c$ as obtained from the mean
(top panel), and from the standard deviation (bottom panel) of the
$j$-th eigenvalue distribution, for $j=1,\ldots, 5$. The purple
filled bands over each $j$ are the combined 1-$\sigma$ estimates of
$\Sigma(j)/\lambda^2$ using the values at different $b$ and $L$,
which are shown using symbols. The unfilled bands in the two panels
enclose all the estimates of $\Sigma(j)/\lambda^2$ at different
$j$, thereby giving an estimate of the systematic error in
$\Sigma/\lambda^2$ due to the large-$N_c$ extrapolations and lattice
spacing effects.}
\eef{cond}

A convenient way to obtain $\Sigma_{\rm lat}(j,N_c,b,L)$ is from
the mean and central moments of the RMM and $\Lambda$ distributions:
\be
\Sigma^{(1)}_{\rm lat}(j,N_c,b,L)\equiv \frac{1}{N_c L^3}\frac{\langle z_j\rangle}{\langle \Lambda_j\rangle};\quad 
\Sigma^{(n)}_{\rm lat}(j,N_c,b,L) \equiv \frac{1}{N_c L^3} \left(\frac{\langle \left(z_j - \langle z_j\rangle\right)^n \rangle}
{\langle \left(\Lambda_j - \langle \Lambda_j\rangle\right)^n \rangle}\right)^{1/n}\quad\text{for $n>1$}.
\label{sigdef}
\ee
If the distributions agree in the large-$N_c$ limit, then the values
of $\Sigma^{(n)}_{\rm lat}(j,N_c,b,L)$ should be the same for all
$n$.  Since, one requires larger statistics to get reliable values
of higher central moments, we restrict ourselves to the mean ($n=1$)
and standard deviation ($n=2$) in this paper.  In \fgn{largen}, we
show the extrapolation of $\Sigma^{(1)}/\lambda^2$ and
$\Sigma^{(2)}/\lambda^2$ to infinite $N_c$ using
$\Sigma(N_c=\infty)/\lambda^2+a_2/N_c+a_3/N_c^2$ ansatz.  It is
clear that the extrapolations of both $\Sigma^{(1)}$ and $\Sigma^{(2)}$
at various $b$, $L$ and $j$ lead to values $\frac{\Sigma}{\lambda^2}\approx
0.004$, significantly away from zero. In \fgn{cond}, we show the
various estimates of $\Sigma/\lambda^2$ (from different $b$, $L$
and five different eigenvalues) in the large-$N_c$ limit.  The top
panel shows the estimates obtained from $\Sigma^{(1)}$ and the
bottom panel for the estimates from $\Sigma^{(2)}$.  It is clear
that $\Sigma/\lambda^2$ from the mean and the standard deviation
of the eigenvalue distributions are consistent with each other. The
estimates of the condensate using the same $j$-th eigenvalue,
$\Sigma(j)/\lambda^2$, at the same lattice spacing but different $L$ are consistent within
errors, thereby serving as a check on continuum reduction which is
a requirement for using smaller $L^3$ lattices. A similar consistency
is also seen between the estimates of $\Sigma(j)/\lambda^2$ at
different lattice spacings, which indicates that our estimate is
close to the continuum value.  Using these independent estimates
of $\Sigma(j)/\lambda^2$ at different $b$ and $L$, we can get a
combined estimate of $\Sigma(j)/\lambda^2$, and we have shown these
as the different purple filled bands superimposed on the data in \fgn{cond}.
We tabulate these values in \tbn{sigtab} for different $j$.  Each
of the tabulated entry is an estimate of the condensate in the
large-$N_c$ limit. It is  evident that these $\Sigma(j)/\lambda^2$,
lie in a narrow range between $0.0038$ and $0.0046$.  Even though
this range of values is small, it is bigger than the statistical
errors in $\Sigma^{(1)}/\lambda^2$.  We take this small variation
in $\Sigma/\lambda^2$ between the eigenvalues to be the systematic
error in our estimate (which could possibly arise due to higher
order $1/N_c$ corrections that we are not able to capture and due
to lattice corrections), and quote our estimate as
\be
\Sigma = (0.0042\pm0.0004) \lambda^2.
\ee
This is shown by the unfilled band in \fgn{cond}.  We checked that
this value is consistent with the estimates from the third central
moments of the eigenvalue distributions, which are noisy compared
to $\Sigma^{(1)}$ and $\Sigma^{(2)}$. Comparing with the value of
string tension, $\sigma$, at $N_c\to\infty$
from~\cite{Karabali:1998yq,Bringoltz:2006zg,Kiskis:2008ah}, we can
express
\be
\frac{\Sigma}{\sigma} = 0.10\pm 0.01.
\ee

The result in this paper implies that SU$(N_c)$ gauge theories
coupled to $2N_f$ flavors of massless fermions must have a confined
phase with a non-zero bilinear condensate.  Our future plan is to
numerically study such theories using massless overlap fermions
with the aim of mapping out the critical line that separates such
a phase from a scale invariant phase.

\bet
\begin{center}
\begin{tabular}{|c||c|c|}
\hline
$j$ & $\Sigma^{(1)}/\lambda^2$ & $\Sigma^{(2)}/\lambda^2$ \\
\hline
1 & 0.0046(2) & 0.0041(1) \\
2 & 0.0042(1) & 0.0039(1) \\
3 & 0.00411(6) & 0.0038(2) \\
4 & 0.00385(6) & 0.0038(2) \\
5 & 0.00383(6) & 0.0038(2) \\
\hline
\end{tabular}
\end{center}
\caption{
Estimates of bilinear condensate obtained using $\Sigma^{(1)}(j)$
and $\Sigma^{(2)}(j)$ from the first five low-lying eigenvalues
$\Lambda_j$ at infinite $N_c$, by a combined fit of the estimates
of $\Sigma^{(1)}(j)$ and $\Sigma^{(2)}(j)$ at different $L$ and
$b$.}
\eet{sigtab}

\acknowledgments
The authors would like to thank Shinsuke Nishigaki and Khandker
Muttalib for discussions on the non-chiral random matrix model used
in this paper.  All computations in this paper were made on the
JLAB computing clusters under a class B project.  The authors
acknowledge partial support by the NSF under grant number PHY-1205396
and PHY-1515446.

\bibliography{biblio}

\begin{thebibliography}{16}
\expandafter\ifx\csname natexlab\endcsname\relax\def\natexlab#1{#1}\fi
\expandafter\ifx\csname bibnamefont\endcsname\relax
  \def\bibnamefont#1{#1}\fi
\expandafter\ifx\csname bibfnamefont\endcsname\relax
  \def\bibfnamefont#1{#1}\fi
\expandafter\ifx\csname citenamefont\endcsname\relax
  \def\citenamefont#1{#1}\fi
\expandafter\ifx\csname url\endcsname\relax
  \def\url#1{\texttt{#1}}\fi
\expandafter\ifx\csname urlprefix\endcsname\relax\def\urlprefix{URL }\fi
\providecommand{\bibinfo}[2]{#2}
\providecommand{\eprint}[2][]{\url{#2}}

\bibitem[{\citenamefont{Karthik and
  Narayanan}(2016{\natexlab{a}})}]{Karthik:2015sgq}
\bibinfo{author}{\bibfnamefont{N.}~\bibnamefont{Karthik}} \bibnamefont{and}
  \bibinfo{author}{\bibfnamefont{R.}~\bibnamefont{Narayanan}},
  \bibinfo{journal}{Phys. Rev.} \textbf{\bibinfo{volume}{D93}},
  \bibinfo{pages}{045020} (\bibinfo{year}{2016}{\natexlab{a}}),
  \eprint{1512.02993}.

\bibitem[{\citenamefont{Karthik and
  Narayanan}(2016{\natexlab{b}})}]{Karthik:2016ppr}
\bibinfo{author}{\bibfnamefont{N.}~\bibnamefont{Karthik}} \bibnamefont{and}
  \bibinfo{author}{\bibfnamefont{R.}~\bibnamefont{Narayanan}}
  (\bibinfo{year}{2016}{\natexlab{b}}), \eprint{1606.04109}.

\bibitem[{\citenamefont{Appelquist and Nash}(1990)}]{Appelquist:1989tc}
\bibinfo{author}{\bibfnamefont{T.}~\bibnamefont{Appelquist}} \bibnamefont{and}
  \bibinfo{author}{\bibfnamefont{D.}~\bibnamefont{Nash}},
  \bibinfo{journal}{Phys. Rev. Lett.} \textbf{\bibinfo{volume}{64}},
  \bibinfo{pages}{721} (\bibinfo{year}{1990}).

\bibitem[{\citenamefont{Damgaard et~al.}(1998)\citenamefont{Damgaard, Heller,
  Krasnitz, and Madsen}}]{Damgaard:1998yv}
\bibinfo{author}{\bibfnamefont{P.~H.} \bibnamefont{Damgaard}},
  \bibinfo{author}{\bibfnamefont{U.~M.} \bibnamefont{Heller}},
  \bibinfo{author}{\bibfnamefont{A.}~\bibnamefont{Krasnitz}}, \bibnamefont{and}
  \bibinfo{author}{\bibfnamefont{T.}~\bibnamefont{Madsen}},
  \bibinfo{journal}{Phys. Lett.} \textbf{\bibinfo{volume}{B440}},
  \bibinfo{pages}{129} (\bibinfo{year}{1998}), \eprint{hep-lat/9803012}.

\bibitem[{\citenamefont{Goldman and Mulligan}(2016)}]{Goldman:2016wwk}
\bibinfo{author}{\bibfnamefont{H.}~\bibnamefont{Goldman}} \bibnamefont{and}
  \bibinfo{author}{\bibfnamefont{M.}~\bibnamefont{Mulligan}}
  (\bibinfo{year}{2016}), \eprint{1606.07067}.

\bibitem[{\citenamefont{'t~Hooft}(1974)}]{'tHooft:1973jz}
\bibinfo{author}{\bibfnamefont{G.}~\bibnamefont{'t~Hooft}},
  \bibinfo{journal}{Nucl.Phys.} \textbf{\bibinfo{volume}{B72}},
  \bibinfo{pages}{461} (\bibinfo{year}{1974}).

\bibitem[{\citenamefont{'t~Hooft}(1983)}]{'tHooft:1983wm}
\bibinfo{author}{\bibfnamefont{G.}~\bibnamefont{'t~Hooft}}, in
  \emph{\bibinfo{booktitle}{{PROGRESS IN GAUGE FIELD THEORY. PROCEEDINGS, NATO
  ADVANCED STUDY INSTITUTE, CARGESE, FRANCE, SEPTEMBER 1-15, 1983}}}
  (\bibinfo{year}{1983}).

\bibitem[{\citenamefont{Narayanan et~al.}(2007)\citenamefont{Narayanan,
  Neuberger, and Reynoso}}]{Narayanan:2007ug}
\bibinfo{author}{\bibfnamefont{R.}~\bibnamefont{Narayanan}},
  \bibinfo{author}{\bibfnamefont{H.}~\bibnamefont{Neuberger}},
  \bibnamefont{and} \bibinfo{author}{\bibfnamefont{F.}~\bibnamefont{Reynoso}},
  \bibinfo{journal}{Phys. Lett.} \textbf{\bibinfo{volume}{B651}},
  \bibinfo{pages}{246} (\bibinfo{year}{2007}), \eprint{0704.2591}.

\bibitem[{\citenamefont{Karabali et~al.}(1998)\citenamefont{Karabali, Kim, and
  Nair}}]{Karabali:1998yq}
\bibinfo{author}{\bibfnamefont{D.}~\bibnamefont{Karabali}},
  \bibinfo{author}{\bibfnamefont{C.-j.} \bibnamefont{Kim}}, \bibnamefont{and}
  \bibinfo{author}{\bibfnamefont{V.~P.} \bibnamefont{Nair}},
  \bibinfo{journal}{Phys. Lett.} \textbf{\bibinfo{volume}{B434}},
  \bibinfo{pages}{103} (\bibinfo{year}{1998}), \eprint{hep-th/9804132}.

\bibitem[{\citenamefont{Bringoltz and Teper}(2007)}]{Bringoltz:2006zg}
\bibinfo{author}{\bibfnamefont{B.}~\bibnamefont{Bringoltz}} \bibnamefont{and}
  \bibinfo{author}{\bibfnamefont{M.}~\bibnamefont{Teper}},
  \bibinfo{journal}{Phys. Lett.} \textbf{\bibinfo{volume}{B645}},
  \bibinfo{pages}{383} (\bibinfo{year}{2007}), \eprint{hep-th/0611286}.

\bibitem[{\citenamefont{Kiskis and Narayanan}(2008)}]{Kiskis:2008ah}
\bibinfo{author}{\bibfnamefont{J.}~\bibnamefont{Kiskis}} \bibnamefont{and}
  \bibinfo{author}{\bibfnamefont{R.}~\bibnamefont{Narayanan}},
  \bibinfo{journal}{JHEP} \textbf{\bibinfo{volume}{09}}, \bibinfo{pages}{080}
  (\bibinfo{year}{2008}), \eprint{0807.1315}.

\bibitem[{\citenamefont{Kiskis et~al.}(2003)\citenamefont{Kiskis, Narayanan,
  and Neuberger}}]{Kiskis:2003rd}
\bibinfo{author}{\bibfnamefont{J.}~\bibnamefont{Kiskis}},
  \bibinfo{author}{\bibfnamefont{R.}~\bibnamefont{Narayanan}},
  \bibnamefont{and}
  \bibinfo{author}{\bibfnamefont{H.}~\bibnamefont{Neuberger}},
  \bibinfo{journal}{Phys. Lett.} \textbf{\bibinfo{volume}{B574}},
  \bibinfo{pages}{65} (\bibinfo{year}{2003}), \eprint{hep-lat/0308033}.

\bibitem[{\citenamefont{Verbaarschot and Zahed}(1994)}]{Verbaarschot:1994ip}
\bibinfo{author}{\bibfnamefont{J.~J.~M.} \bibnamefont{Verbaarschot}}
  \bibnamefont{and} \bibinfo{author}{\bibfnamefont{I.}~\bibnamefont{Zahed}},
  \bibinfo{journal}{Phys. Rev. Lett.} \textbf{\bibinfo{volume}{73}},
  \bibinfo{pages}{2288} (\bibinfo{year}{1994}), \eprint{hep-th/9405005}.

\bibitem[{\citenamefont{Szabo}(2001)}]{Szabo:2000qq}
\bibinfo{author}{\bibfnamefont{R.~J.} \bibnamefont{Szabo}},
  \bibinfo{journal}{Nucl. Phys.} \textbf{\bibinfo{volume}{B598}},
  \bibinfo{pages}{309} (\bibinfo{year}{2001}), \eprint{hep-th/0009237}.

\bibitem[{\citenamefont{Mehta}(2004)}]{Mehta:2004}
\bibinfo{author}{\bibfnamefont{M.~L.} \bibnamefont{Mehta}},
  \bibinfo{journal}{Random Matrices, Pure and Applied Mathematices Series}
  \textbf{\bibinfo{volume}{142}} (\bibinfo{year}{2004}).

\bibitem[{\citenamefont{Nishigaki}(2016)}]{Nishigaki:2016nka}
\bibinfo{author}{\bibfnamefont{S.~M.} \bibnamefont{Nishigaki}}, in
  \emph{\bibinfo{booktitle}{{Proceedings, 33rd International Symposium on
  Lattice Field Theory (Lattice 2015)}}} (\bibinfo{year}{2016}),
  \eprint{1606.00276},
  \urlprefix\url{https://inspirehep.net/record/1466628/files/arXiv:1606.00276.pdf}.

\end{thebibliography}
\end{document}